\documentclass[superscriptaddress,aps,preprintnumbers,amsmath,showpacs,amssymb,prd,nofootinbib,reprint]{revtex4-1}
\usepackage{bm, color} 
 \usepackage{amssymb,amsfonts,slashed,amsthm,amsmath,graphicx, soul}

\usepackage{footmisc}
\usepackage{hyperref}
\begin{document}

\renewcommand{\thesection}{\Alph{section}}

\newcommand{\mplank}{\textrm{M}_{\textrm{P}}}
\newcommand{\mg}{m_{\gamma^{\prime}}}
\newcommand{\higgs}{H_{\scriptsize \rm h}}
\newcommand{\higgst}{\tilde{H}_{\scriptsize \rm h}}
\renewcommand{\Re}{\mathrm{Re}}
\newcommand{\MF}{{\sf B}}

\newcommand{\muu}{m_{\gamma^{\prime}}}
\newcommand{\chie}{\chi_{\rm eff}}
\newcommand{\ve}[2]{\left(
\begin{array}{c}
 #1\\
#2
\end{array}
\right)
}

\renewcommand\({\left(}
\renewcommand\){\right)}
\renewcommand\[{\left[}
\renewcommand\]{\right]}

\def\ebq{\end{equation} \begin{equation}}
\renewcommand{\figurename}{Figure.}
\renewcommand{\tablename}{Table.}
\newcommand{\Slash}[1]{{\ooalign{\hfil#1\hfil\crcr\raise.167ex\hbox{/}}}}
\newcommand{\bra}[1]{ \langle {#1} | }
\newcommand{\ket}[1]{ | {#1} \rangle }
\newcommand{\beq}{\begin{equation}}  \newcommand{\eeq}{\end{equation}}
\newcommand{\bef}{\begin{figure}}  \newcommand{\eef}{\end{figure}}
\newcommand{\bec}{\begin{center}}  \newcommand{\eec}{\end{center}}
\newcommand{\non}{\nonumber}  \newcommand{\eqn}[1]{\begin{equation} {#1}\end{equation}}
\newcommand{\laq}[1]{\label{eq:#1}}  
\newcommand{\dd}[1]{{d \o d{#1}}}
\newcommand{\Eq}[1]{Eq.~(\ref{eq:#1})}
\newcommand{\Eqs}[1]{Eqs.~(\ref{eq:#1})}
\newcommand{\eq}[1]{(\ref{eq:#1})}
\newcommand{\Sec}[1]{Sec.\ref{chap:#1}}
\newcommand{\ab}[1]{\left|{#1}\right|}
\newcommand{\vev}[1]{ \left\langle {#1} \right\rangle }
\newcommand{\bs}[1]{ {\boldsymbol {#1}} }
\newcommand{\lac}[1]{\label{chap:#1}}
\newcommand{\SU}[1]{{\rm SU{#1} } }
\newcommand{\SO}[1]{{\rm SO{#1}} }

\def\({\left(}
\def\){\right)}
\def\dt{{d \o dt}}
\def\diag{\mathop{\rm diag}\nolimits}
\def\Spin{\mathop{\rm Spin}}
\def\O{\mathcal{O}}
\def\U{\mathop{\rm U}}
\def\Sp{\mathop{\rm Sp}}
\def\SL{\mathop{\rm SL}}
\def\tr{\mathop{\rm tr}}
\newcommand{\OR}{~{\rm or}~}
\newcommand{\AND}{~{\rm and}~}
\newcommand{\EV}{ {\rm \, eV} }
\newcommand{\KEV}{ {\rm \, keV} }
\newcommand{\MEV}{ {\rm \, MeV} }
\newcommand{\GEV}{ {\rm \, GeV} }
\newcommand{\TEV}{ {\rm \, TeV} }

\def\o{\over}
\def\a{\alpha}
\def\b{\beta}
\def\c{\varepsilon}
\def\d{\delta}
\def\e{\epsilon}
\def\f{\phi}
\def\g{\gamma}
\def\h{\theta}
\def\k{\kappa}
\def\l{\lambda}
\def\m{\mu}
\def\n{\nu}
\def\p{\psi}
\def\q{\partial}
\def\r{\rho}
\def\s{\sigma}
\def\t{\tau}
\def\u{\upsilon}
\def\v{\varphi}
\def\w{\omega}
\def\x{\xi}
\def\y{\eta}
\def\z{\zeta}
\def\D{\Delta}
\def\G{\Gamma}
\def\H{\Theta}
\def\L{\Lambda}
\def\F{\Phi}
\def\P{\Psi}
\def\S{\Sigma}
\def\me{\mathrm e}
\def\ol{\overline}
\def\tl{\tilde}
\def\*{\dagger}

\newcommand{\exclude}[1]{}

\def\bra{\langle}
\def\ket{\rangle}
\def\beq{\begin{equation}}
\def\eeq{\end{equation}}
\newcommand{\C}[1]{\mathcal{#1}}
\def\ov{\overline}

\preprint{TU-1173}

\title{ 
Weak-Scale Higgs Inflation 
}

\author{Wen Yin}
\affiliation{Department of Physics, Tohoku University, Sendai, Miyagi 980-8578, Japan }

\begin{abstract}
The present measurement of the standard model (SM) parameters suggests that the Higgs effective potential has a maximum at the intermediate scale, and the electroweak (EW) vacuum is not absolutely stable. 
The simplest possibility for absolute EW stability may be introducing a very large Higgs-Ricci scalar non-minimal coupling. 
In this extension of the SM, I study the cosmic inflation driven by the Higgs field. Since the resulting Hubble parameter happens to be around the weak scale, I call this scenario {\it weak-scale Higgs inflation}.
It is pointed out that the cosmic-microwave background normalization of the scalar density perturbation can be enhanced compared with the conventional Higgs inflation prediction, thanks to the potential maximum predicted by the SM.
It turns out in the Palatini formulation of the gravity with various higher dimensional terms suppressed by the scale close to the Planck scale that the inflation is successful, with a prediction of the running of the spectral index, $\a_s=-(3-4)\times 10^{-3}$, which can be tested in the future. I also argue the UV (in)sensitivity of the prediction, the fact that the parameter region for successful inflation is very close to the criticality for potential shapes for eternal and non-eternal inflation, and the phenomenological applications. 

\noindent
\end{abstract}
\maketitle
\flushbottom

\section{Introduction}
Cosmic inflation~\cite{Starobinsky:1980te,Guth:1980zm,Sato:1980yn,Linde:1981mu,Albrecht:1982wi}, which generates the primordial density perturbation, is the key assumption of the modern inflationary cosmology, which is strongly suggested from the recent cosmic-microwave background (CMB) data~\cite{Planck:2018jri,Planck:2018vyg}. From the field-theoretical point of view, there should be a scalar field, called inflaton, with a very flat potential to drive slow-roll inflation. The potential should not be completely flat for the inflation to end, and then the inflaton decays to reheat the Universe, connecting to the big-bang cosmology. The particle origin of the inflaton is a leading mystery of cosmology. 

The Higgs inflation~\cite{Bezrukov:2007ep,Bezrukov:2008ej} (see also \cite{Rubio:2018ogq}),  in which the Higgs boson is the inflaton, was considered as the minimal possibility for the inflation to explain the primordial density perturbation of the Universe. 
 In the scenario, a large Higgs-Ricci scalar non-minimal coupling, $\x$,  makes Higgs potential flat for the large Higgs field. 
 Although the recent measurement of the SM disfavors the minimal scenario, 
the standard model (SM) coupling may be significantly different when the Higgs field is large because model violates the perturbative unitarity \cite{Bauer:2010jg, Bezrukov:2014ipa}. This is for the metric formulation of gravity because the kinetic term of the Higgs boson is enhanced in the Einstein frame when the Higgs field is slightly below the slow-roll regime. 
By introducing certain smooth transitions of the SM couplings to the arbitrary value at the high energy scale, hilltop/inflection-point-type Higgs inflation was considered in \cite{Fumagalli:2016lls,Rasanen:2017ivk,Enckell:2018kkc}. 
In the Palatini formulation of gravity, in which the Levi-Civita connection and metric are independent geometrodynamical variables,
 the unitarity violation is not severe~\cite{Bauer:2010jg}, while the spectral index is sensitive to the higher dimensional terms~\cite{Jinno:2019und}. 
In the previous studies, the authors have focussed on the relatively high inflation scale. The reason why the Higgs inflation with an almost tree-level quartic coupling cannot have successful inflation in the regime of intermediate potential energy is that the change of the potential slope is too slow to have a large enough power spectrum of the scalar curvature perturbation at the horizon exit of the CMB scale. This situation cannot be fully resolved by introducing higher dimensional operators~\cite{Jinno:2019und, Gialamas:2019nly, Gialamas:2020vto}.

The precision measurement of the SM parameters, on the other hand, suggests that the Higgs quartic coupling, $\l$, turns to negative at an instability scale, (see Refs.\,\cite{Workman:2022ynf} and, e.g., \cite{Li:2022ugn})
\beq \L_I=10^{9-12}\GEV,\eeq 
at which $\l(\L_I)=0$, which disfavors the minimal high scale Higgs inflation without changing the SM couplings. 
In addition, although the electroweak (EW) false vacuum has a lifetime above the age of the Universe within the quantum field theory~\cite{Sher:1988mj,Arnold:1989cb,Anderson:1990aa,Arnold:1991cv,Espinosa:1995se,Isidori:2001bm,Espinosa:2007qp,Ellis:2009tp,EliasMiro:2011aa,Bezrukov:2012sa,Degrassi:2012ry,Buttazzo:2013uya,Bednyakov:2015sca,Andreassen:2017rzq,Chigusa:2017dux,Chigusa:2018uuj}, the inflationary fluctuation~\cite{Kobakhidze:2013tn}, the preheating after inflation~\cite{Herranen:2015ima,Ema:2016kpf,Li:2022ugn}, right-handed neutrinos~\cite{Chigusa:2018uuj},
the present Universe's small blackholes~\cite{Hiscock:1987hn, Berezin:1987ea} and compact objects without a horizon \cite{Oshita:2018ptr}, and even the pure gravitational effect~\cite{Oshita:2021aux} may let it decay into the true vacuum within the age of the Univese. Perhaps the EW vacuum needs to be a true one, which suggests some beyond SM physics e.g. \cite{Lebedev:2012zw, Elias-Miro:2012eoi,Nakayama:2021avl, Matsui:2020wfx}.

Although it was not clearly pointed out in the literature, the simplest possibility to make the EW vacuum absolutely stable should be introducing a very large non-minimal coupling, $\x$ (\Sec{1}), as done in the Higgs inflation.\footnote{See c.f. Ref.\,\cite{Czerwinska:2016fky} for the study of the suppression of the vacuum decay rate with the non-minimal coupling.}  This makes the potential flat before the Higgs quartic coupling runs to a negative value. 
Thus absolute stability is obtained without introducing UV corrections to the potential. 

In this paper, I will show that the same setup can lead to a Higgs inflation with $\L_I$ scale potential with consistent CMB normalization if the non-minimal coupling is properly chosen thanks to the Higgs hilltop predicted by the SM (\Sec{2}). This happens even without introducing UV corrections to the SM parameters. The $\L_I$ potential energy corresponds to a weak-scale Hubble parameter, and thus, I call it {\it weak-scale Higgs inflation}. 
In a Palatini formulation of the Higgs inflation, I find that explaining the spectral index with generic Planck suppressed term~\cite{Jinno:2019und} universally predicts the sizable running of the spectral index that can be fully tested in the CMB-S4~\cite{CMB-S4:2016ple} and SPHEREx~\cite{Dore:2014cca} together with DESI~\cite{DESI:2013agm}, WFIRST~\cite{https://doi.org/10.48550/arxiv.1503.03757}, or SKA~\cite{CosmologySWG:2015ysq} (\Sec{3}). 
This prediction is robust because various higher dimensional terms and higher loop effects can be interpreted as the redefinition of an effective potential with few relevant terms during inflation. 
The potential for successful inflation happens to lie very close to the critical parameter regime between the eternal inflation and non-eternal inflation to take place (\Sec{4}). Phenomenological applications are also argued.

\section{A simple possibility for the electroweak vacuum stability.}
\lac{1}

Let me introduce the setup, together with pointing out the
 simple (perhaps simplest) possibility to make the EW vacuum absolutely stable. This is to introduce a very large non-minimal coupling, $\x$, in the Jordan frame
\beq
\laq{nonmini}
 {\cal L }\supset -\frac{1}{2}M_{\rm pl}^2\Omega^2g^{\m\n}{\cal R}_{\m\n}
\eeq 
with ${\cal R}_{\m\n}$ being the Ricci curvature tensor, 
\beq \Omega\equiv \(1+\x \frac{ h^2}{M_{\rm pl}^2}\)^{1/2},\eeq
$h$ the neutral component of the Higgs field, we take $M_{\rm pl}= 2.4\times 10^{18}\GEV$ for the reduced Planck scale, $g_{\m\n}$ the metric. 
The action is $S=\int d^4x\sqrt{|\det{g}|} {\cal L}$.
Then moving to the Einstein frame via a Weyl transformation, we obtain the 
Higgs 1PI potential 
\beq
\laq{1PI}
V_{EF}[h,\m_H^2]=V_{JF}\[\frac{{h}}{\Omega},\frac{\m_H^2}{\Omega^2}\],
\eeq
with $\m_H^2 (>0)$ being the Higgs mass parameter. Here and hereafter, $X_{EF} \AND X_{JF}$ represent the quantity, $X$, in the Einstein and Jordan frame, respectively. 

Let us take the renormalization scale to be $\L_I$ so that the tree-level quartic coupling vanishes. 
Then  the Jordan frame 1PI effective potential is
\beq
\laq{eft}
V_{JF}\approx -\frac{3y_t^4 h^4}{64\pi^2}\log[\frac{y_t^2 h^2}{2\L_I^2}] -\m_H^2 \frac{h^2}{2},
\eeq
where $y_t[\L_I]\simeq 0.5$ is the top Yukawa coupling. We omit the various irrelevant terms,\footnote{We may include contributions from other SM couplings to the $y_t$ term. The leading log contributions can be written again in the same form by slightly redefining $\L_I,y_t$. 
The contribution from higher order terms for the SM Higgs potential (see \cite{Degrassi:2012ry}) will not change our conclusions since they will not change the form of \Eqs{VEF} and \eq{dEF}.
} e.g., the term for canceling the cosmological constant.  
This potential has a hilltop at 
\beq
h= h_{\rm hilltop} \equiv \sqrt{e^{-1/2}\frac{2 \L^2_I}{y_t^2}}
\eeq
at which $V_{JF}'\simeq 0$ by neglecting the tiny $\m_H^2$ term.

\Eq{1PI} applies  at the quantum level, e.g., the top mass is changed from $y_t h/\sqrt{2}\to y_t h/(\sqrt{2}\Omega)$ affecting the Coleman-Weinberg potential, the potential becomes flat with 
$ h \gtrsim \L_{\rm flat}\equiv \frac{M_{\rm pl}}{\sqrt{\x}}.
$
If \beq \laq{cond} \L_{\rm flat}\lesssim \L_I \to \x\gtrsim M_{\rm pl}^2/\L_I^2\eeq
the EW vacuum in the Einstein frame is the true one.

Let me comment on, in general, that $h$ is not the canonically normalized field. 
The canonically normalized field, $\f$, can be obtained from the wave function of $h$ in the Einstein frame,
\beq
{\cal L}_{\rm kin}=  Z[h] |\partial_\mu h|^2,
\eeq
via the relation 
$
\frac{d \f}{d h}= Z[h]^{1/2}.
$
Thus we should understand $h$ in $V_{EF}$ as a function of $\phi$ determined by this relation. 
The field redefinition does not change the potential value. 
Also, since $h$ is usually a monotonically increasing function of $\f$,  the existence or the number of the vacua is the same for $h$ and $\f$.
In particular, our discussions except for those in \Sec{3} apply to both  the metric formulation ($Z=\Omega^{-2} + \frac{3}{2}(M_{\rm pl}d\log(\Omega^2)/d h)^2$) and the Palatini formulation ($Z=\Omega^{-2}$).

Two examples (Palatini formulation and metric formulation) of the Higgs potential satisfying the condition \eq{cond} are given in Fig.\ref{fig:1}. 
\begin{figure}[t!]
\begin{center}  
\includegraphics[width=80mm]{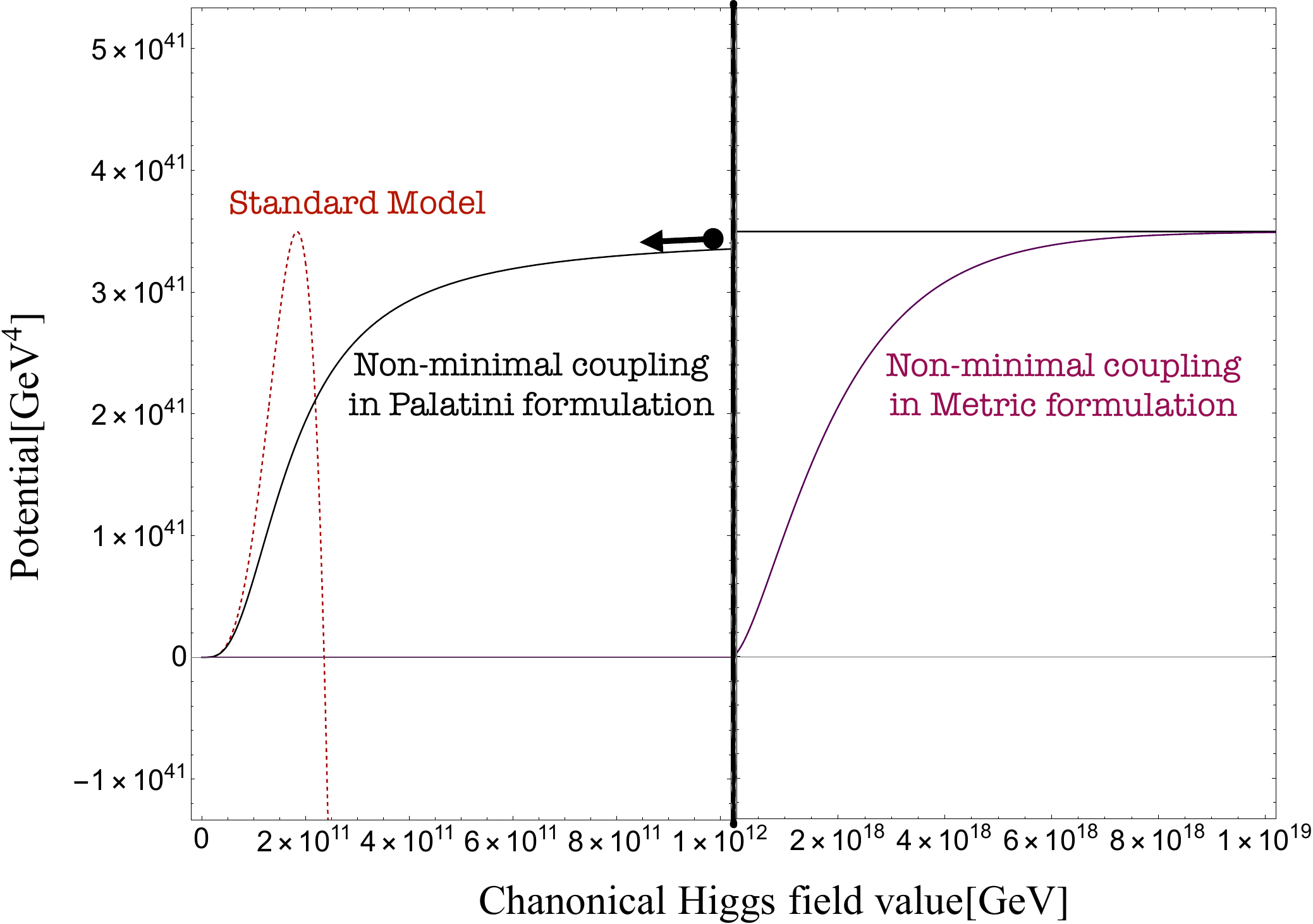}
\end{center}
\caption{The Higgs potential with non-minimal coupling in the Palatini formalism (black solid line) and metric formalism (Purple solid line) from top to bottom. Here we take $\x=\sqrt{e} y_t^2\frac{ M_{\rm pl}^2}{2 \Lambda_I^2}$ (see the discussion for inflation). 
The red dashed line is the SM one without a non-minimal coupling, in which the EW vacuum is not the absolute minimum. We fix $\Lambda_I=10^{11}\GEV,y_t=0.5, \m_H^2=\(125\GEV\)^2/2.$ 
}
\label{fig:1}
\end{figure}

Before ending this part, I will comment on the perturbative unitarity of the scenario. In the metric formalism, we can see that kinetic term with $h \ll M_{\rm pl}/\sqrt{\x}$ has a contribution of $Z (\partial h)^2\supset 6\x^2 \frac{h^2}{M_{\rm pl}^2} (\partial h)^2$ and the perturbative unitarity may be violated at $M_{\rm pl}/\x \lesssim 2\TEV\(\frac{\L_I}{10^{11}\GEV}\)^2$.\footnote{It might be interesting to relate this cutoff scale to the weak scale.} In the Palatini formulation we do not have this sizable kinetic term, and the perturbative unitarity is valid up to $\sim M_{\rm pl}/\sqrt{\x}\sim \L_I$. During the inflation, the cutoff, which is $h$ dependent, is even higher~\cite{Bauer:2010jg}.

\section{Weak-scale Inflation by Higgs potential.}
\lac{2}

The flat potential for absolute EW vacuum stability may drive cosmic inflation.
For slow-roll inflation to occur, we need both  the slow-roll parameters to satisfy $|\varepsilon|, |\eta|\ll 1$, where 
$\varepsilon(\phi) \equiv \frac{M_{\rm pl}^2}{2} 
\left(\frac{V_{EF,\f}}{V_{EF}}\right)^2= \frac{M_{\rm pl}^2}{2} \left(\frac{V_{EF,h}}{V_{EF}}\right)^2 
Z^{-1}
$
and 
$
\eta({\f}) \equiv M_{\rm pl}^2 \frac{V_{EF,\f\f}}{V_{EF}}= M_{\rm pl}^2 \frac{ Z^{-1/2}d (Z^{-1/2} V_{EF,h})/dh}{V_{EF}}.
$
In this paper, $X_{,Y}$ means $dX/dY$. 
It is obvious that without a non-minimal coupling, the slow-roll condition cannot be satisfied due to the too-large SM couplings and the induced Coleman-Weinberg corrections.
Thus I will focus on the parameter region of the hilltop at $h_{\rm hilltop}>M_{\rm pl}/\sqrt{\x} $. 
It is then convenient to use the parameterization,
\beq\laq{xi}
\xi=\x_0+ \d \xi, ~\x_0= \sqrt{e} y_t^2\frac{M_{\rm pl}^2 }{2 \Lambda_I^2}.
\eeq
When $\d \xi=0$ and $\mu_H^2=0$, we find $\lim_{h\to \infty} \frac{h}{\Omega} \to h_{\rm hilltop}
$. 
By expanding in $\L_I/h, \d \x/\x_0, \AND \m_H^2/\L^2_I$ we get 
\begin{align} 
\laq{VEF}
V_{EF}&\approx V_0\(1+  \k \L_I^2 h^{-2} 
-\frac{8}{ e y_t^4} \Lambda_I^4 h^{-4}+ \O(\L_I^6 h^{-6})\),
\end{align} 
where
\beq
V_0\approx \frac{3\L_I^4}{32e \pi^2}, \k \approx -\frac{64 \pi ^2}{3 y_t^4} \frac{\left(\m_H^2+24  H_{\rm inf}^2 \delta\x \right)}{\L_I^2}, H^2_{\rm inf}\equiv \frac{V_0}{3M_{\rm pl}^2}. 
\laq{kappa}
\eeq
Here we only show the leading order terms in $\d \x/\x_0, \m_H^2$ in the coefficient for each order of $1/h$. 
Notice that $\mu_H^2$ is comparable to the inflationary Hubble parameter, $H_{\rm inf}^2$, and the Higgs mass term cannot be neglected in the inflationary dynamics.\footnote{The QCD scale is also around $\m_H$ due to the heavy quarks during inflation. Their contributions is at most $V_{QCDEF}\propto y_t^2/(\x^2 h^3) $ via the Higgs-top coupling at large $h$ and is neglected compared with the Higgs mass term. The term to cancel the vacuum energy is more suppressed. }

From the slow-roll equation, $3H \dot\f\approx -V_{{EF}, \f},$ the e-folding number can be obtained as $N\approx \int^{\f_*}_{\f_{\rm end}}{\(\frac{V_{{EF},\f}}{3H^2}\)^{-1} d\f }\approx \int^{h_*}_{h_{\rm end}}{\({\frac{V_{ EF}}{V_{EF,h} M_{\rm pl}^2}}\) Z dh },$ with $H\approx \sqrt{V_{EF}/(3 M_{\rm pl}^2)}$ being the Hubble parameter, and $\dot{X}\equiv d X/dt$, $t$ the cosmic time. 
$\f_{\rm end}~(h_{\rm end})$ is the field value that one of the slow conditions is violated, i.e., when $\max{[|\varepsilon(\f_{\rm end})|, |\eta(\f_{\rm end})|]}=1.$ At the horizon exit $\f=\f_*$ corresponds to the CMB pivot scale, where the power spectrum of the curvature perturbation, the spectral index, and its runnings are measured. Here and hereafter, the subscript $*$ denotes the quantity at the horizon exit.
The e-folding number should be matched with the thermal history. Soon after the inflation, reheating should complete due to the fast Higgs interaction rate, and we get the radiation-dominated Universe immediately. 
This implies 
\beq
N\approx 43 +\log[\frac{\Lambda_I}{10^{11}\GEV}] ,
\eeq
which is smaller than the usual Higgs inflation. 

The power spectrum of the curvature perturbation is related to the inflaton potential shape as 
\begin{equation}
\laq{pln}
\Delta^2_{\mathcal{R}}(k) \simeq \(\frac{H^2}{2\pi \dot{\phi}}\)^2 \simeq 
\frac{V_{EF}(\phi)^3}{12 \pi^2 V_{EF}'(\phi)^2 M_{\rm pl}^6},
\end{equation} 
by matching the CMB data. 
For instance, the measured CMB normalization~\cite{Planck:2018jri,Planck:2018vyg} 
\begin{equation}
\label{pn}
\Delta^{2}_{\mathcal{R}, \rm CMB}(k_*) \simeq 2.1
\times 10^{-9},
\end{equation}
at the pivot scale $k_* = 0.05\, {\rm Mpc}^{-1}$ gives a relation between the potential hight and  the derivative at the horizon exit. 
In the usual Higgs inflation with a  low scale Hubble parameter, the CMB normalization is very difficult to satisfy because only the $\O(h^{-2})$ term will contribute to the slow roll around the horizon exit. This would lead to a very tiny contribution to $\Delta_{\cal R}^2[\f_*]\simeq 
\frac{\xi  N^2 {V_0}}{3 \pi^2 M_{\rm pl}^4} \AND \frac{N^2 {V_0}}{18 \pi ^2 M_{\rm pl}^4} $ for Palatini and metric formulations, respectively, since $V_0\ll M_{\rm pl}^4. $
In our scenario,  $|\k|$ can be much smaller than the conventional value, and thus at not very large $h$, both $h^{-2}$ and $h^{-4}$ terms are important for the classical motion of the inflaton field.  
In particular, with a certain positive $\k$, in the finite range of $h$, there can be a hilltop, which corresponds to the hilltop within the SM. 
Then, the horizon exit becomes around the hilltop with suppressed $\varepsilon$ and enhanced $\Delta_{\cal R}^2$, explaining the CMB normalization. 
This is shown in Fig.~\ref{fig:2}, where I
display $\Delta_{\cal R}^2$ by varying $\d \x$ in \Eq{VEF}. I take $y_t=0.5, \m_H^2=(125\GEV)^2/2$ and $\L_I=10^{9,10,11,12}\GEV$ from top to bottom.\footnote{Strictly speaking, $y_t$ relates to $\L_I$, and  $y_t$ also changes less than $10\%$ with the choices of $\L_I$. A more precise analysis by including various subdominant corrections, including the $\ddot\phi$ term in the equation of motion, will be given elsewhere. } 
The enhancement of $\Delta_{\cal R}^2$ to the desired value occurs in the Palatini and metric formulations.

\begin{figure}[t!]
\begin{center}  
\includegraphics[width=80mm]{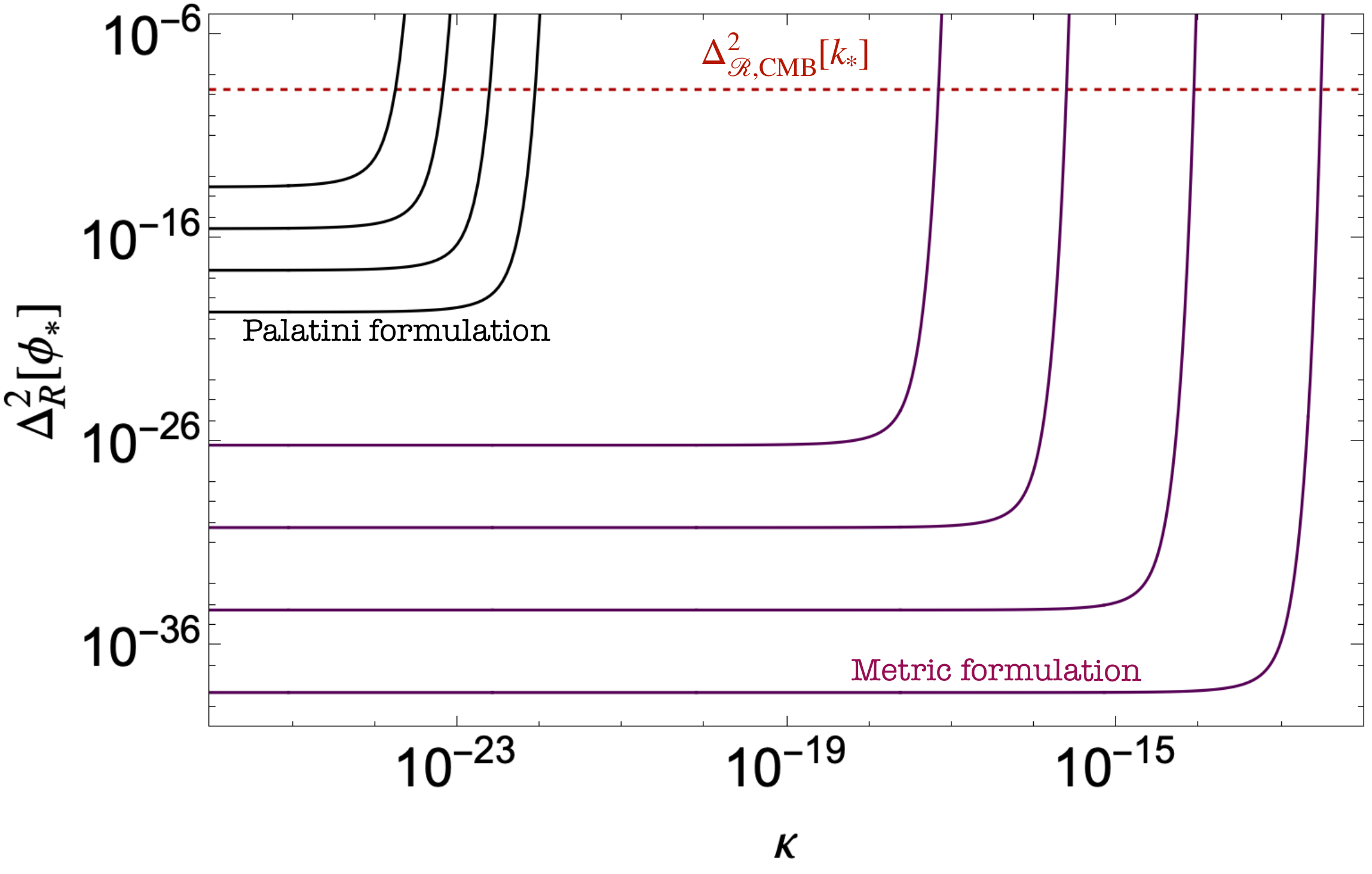}
\end{center}
\caption{ 
The CMB normalization by varying $\k$ in \Eq{VEF}. 
$\Lambda_I=10^9,10^{10},10^{11},10^{12}\GEV$ from top to bottom with fixed $y_t=0.5$ and $\mu_H^2=\(125\GEV\)^2/2$ 
}
\label{fig:2}
\end{figure}

\section{  Spectral indices and UV (in)sensitivity in weak-scale Palatini Higgs inflation }
\lac{3}
With the SM prediction, $\L_I=10^{9-12}\GEV$, however, we found the predicted scalar spectral index 
$n_s \simeq  1- 6 \varepsilon(\f_*) + 2\eta (\f_{*}),$ 
 is too small. For instance, $n_s\sim 0.6$ for the Palatini formalism with $\L_I=10^{11}\GEV$.\footnote{For the Palatini (metric) formulation with $\L_I\simeq 10^{14}\GEV, (10^{16}\GEV)$ we can have the spectral index to be consistent with the observation. This is in tension with the SM prediction. } 
This should be compared with the measured one $n_s^{\rm CMB}=0.9665\pm 0.0038$ \cite{Planck:2018jri,Planck:2018vyg}.

$n_s$ in the Palatini formulation is UV sensitive~\cite{Jinno:2019und}, i.e., adding Planck scale-suppressed terms would change the prediction of $n_s$~(see also new inflation \cite{Takahashi:2013cxa}, multi-natural inflation\cite{Czerny:2014wza, Czerny:2014xja,Czerny:2014qqa,Higaki:2014sja, Croon:2014dma,Higaki:2015kta, Higaki:2016ydn} and very low scale inflation by ALP~\cite{Daido:2017wwb, Daido:2017tbr,  IAXO:2019mpb,Takahashi:2019qmh,Marsh:2019bjr, Takahashi:2021tff} with a similar mechanism to enhance $n_s$.). In the metric formulation, to enhance $n_s$ we need much stronger higher dimensional terms than the Planck scale and, moreover, there may be issues on the perturbative unitarity~\cite{Bauer:2010jg,Bezrukov:2014ipa}.  
 Thus we will focus on the Palatini formulation with higher dimensional terms suppressed by the Planck scale to demonstrate the consistency with the CMB data.  

By concentrating the field value in the range $M_{\rm pl}/\sqrt{\x}\lesssim h \ll M_{\rm pl}$, where inflation takes place, the higher dimensional term can be understood in the effective theory. 
In the Einstein frame, in a class of UV models, we will get \beq
\laq{dEF}
\d V_{EF}\approx \frac{V_0}{M^2} h^2 
\eeq
as the leading correction.
For instance, this is the case of a dimension six term, $3y_t^4 h^6/(128\pi^2 M^2)$ in the Jordan frame Higgs potential~\cite{Jinno:2019und}. 
This can also be obtained with corrections to the couplings. The corrected bottom quark Yukawa coupling \beq 
\laq{Yukawa}
{\cal L}_{JF}\supset -y_b (1-\frac{y_t^4 h^2}{4 y_b^4 M^2}) \frac{h}{\sqrt{2}} b_L b_R\eeq leads to the desired form with a logarithmic correction.  
I will explain why a similar correction in the top Yukawa interaction is suppressed in the next paragraph. 

There are also irrelevant UV corrections. They include the modification of $\Omega$. 
To see this, let us remind that in low-scale inflation, the first slow-roll parameter $\varepsilon$ is usually highly suppressed.
Thus $|\partial V_{EF}/\partial \Omega |=  h\Omega |V_{EF,h}|+ \O(\m_H^2)= \sqrt{2\varepsilon} V_{EF}  h/M_{\rm pl} +\O(\m_H^2) $ is also suppressed. Here we used $Z^{-1/2} = \Omega$ for the Palatini formalism. 
A Planck suppressed operator in $\Omega^2$ in \Eq{nonmini} rarely affects the prediction.\footnote{The smooth and monotonic wave function $Z$ gets a correction $d\log Z/d(\Omega^2)\times \O(h_{\rm hilltop}^2/M_{\rm pl}^2)$. It is negligible in the estimation of $N[\f_*], \Delta_{\cal R}^2[\f_*] \AND n_s[\f_*]$.} 
A similar correction of \Eq{Yukawa} to the top Yukawa interaction is also suppressed. This is because 
by taking other couplings to zero and $\d \x\to0$,
 it can be seen as the redefinition of $\Omega.$ 
Alternatively, the running effect may induce a correction to $\Omega$,~\cite{Bezrukov:2009db} $\d (\Omega^2) 
 \sim\frac{ 3\xi y_t^2 }{4\pi^2} \frac{h^2}{M_{\rm pl}^2}\log{[\frac{y_t^2h^2}{2\L_I^2(1+\x h^2/M^2_{\rm pl})}]}= -\frac{3 \sqrt{e}  y_t^4}{16 \pi ^2 \Lambda_I^2}h^2-\frac{3 y_t^2}{4 \pi ^2}
+\frac{3 \Lambda_I^2}{4 \sqrt{e} \pi^2 }h^{-2}
 +\O(h^{-4},\delta \x/\x_0)$ at large $h.$ 
The first two terms redefine the original $\Omega$, while the $\O(h^{-2})$ term contributions are highly suppressed due to the aforementioned insensitivity. They also do not affect the final results. 

\begin{figure}[t!]
\begin{center}  
\includegraphics[width=60mm]{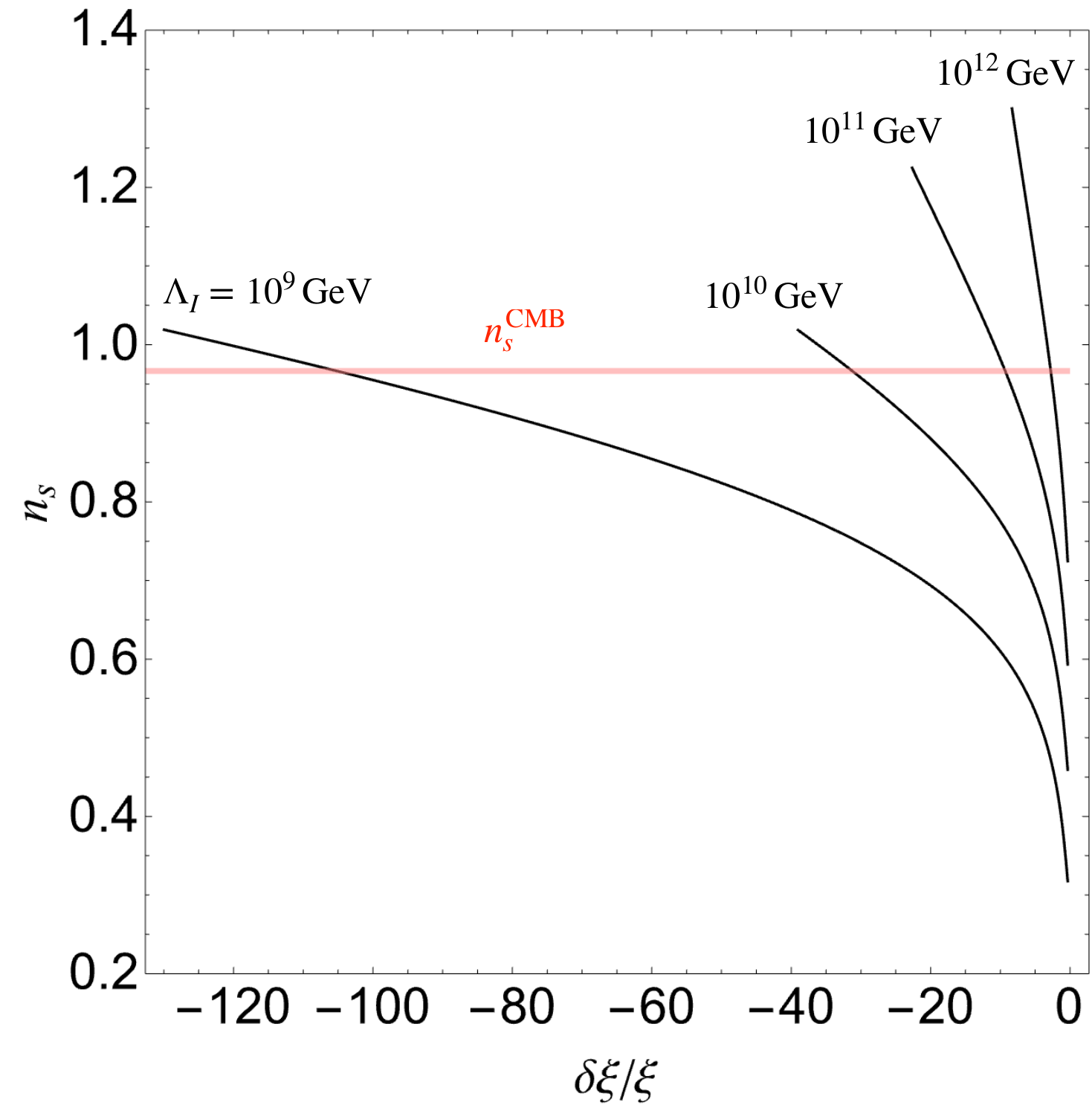}
\includegraphics[width=65mm]{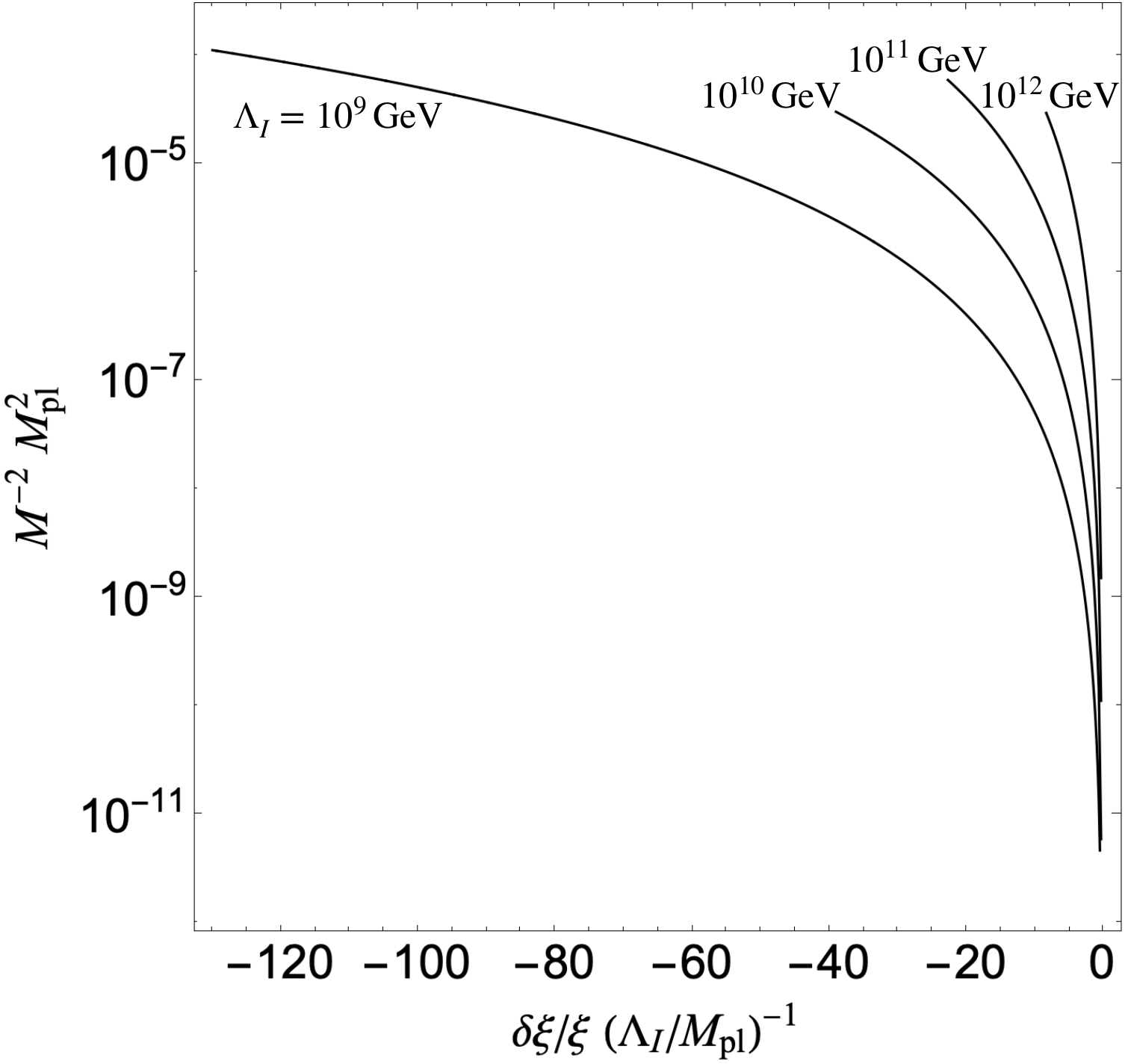}
\end{center}
\caption{ 
$n_s$ vs $\d \x/\x_0$ [upper panel], and $M^{-2} M_{\rm pl}^2$ vs $\d \x/\x_0$ [lower paner]. 
We take $\Delta_{\cal R}^2[\f_*]= \Delta^2_{{\cal R},{\rm CMB}}[k_*]$ by varying the one-dimensional combination of $\d \x\AND M^{-2}$. 
Notice that the normalization of $\L_I/M_{\rm pl}$ are different for lines with different $\L_I$.  
The dependency of $M^{-2} M_{\rm pl}^3/\L_I$ on $\d\x/\x_0 M_{\rm pl}/\L_I $ is almost irrelevant to $ \L_I$. 
}
\label{fig:4}
\end{figure}

\begin{figure}[t!]
\begin{center}  
\includegraphics[width=65mm]{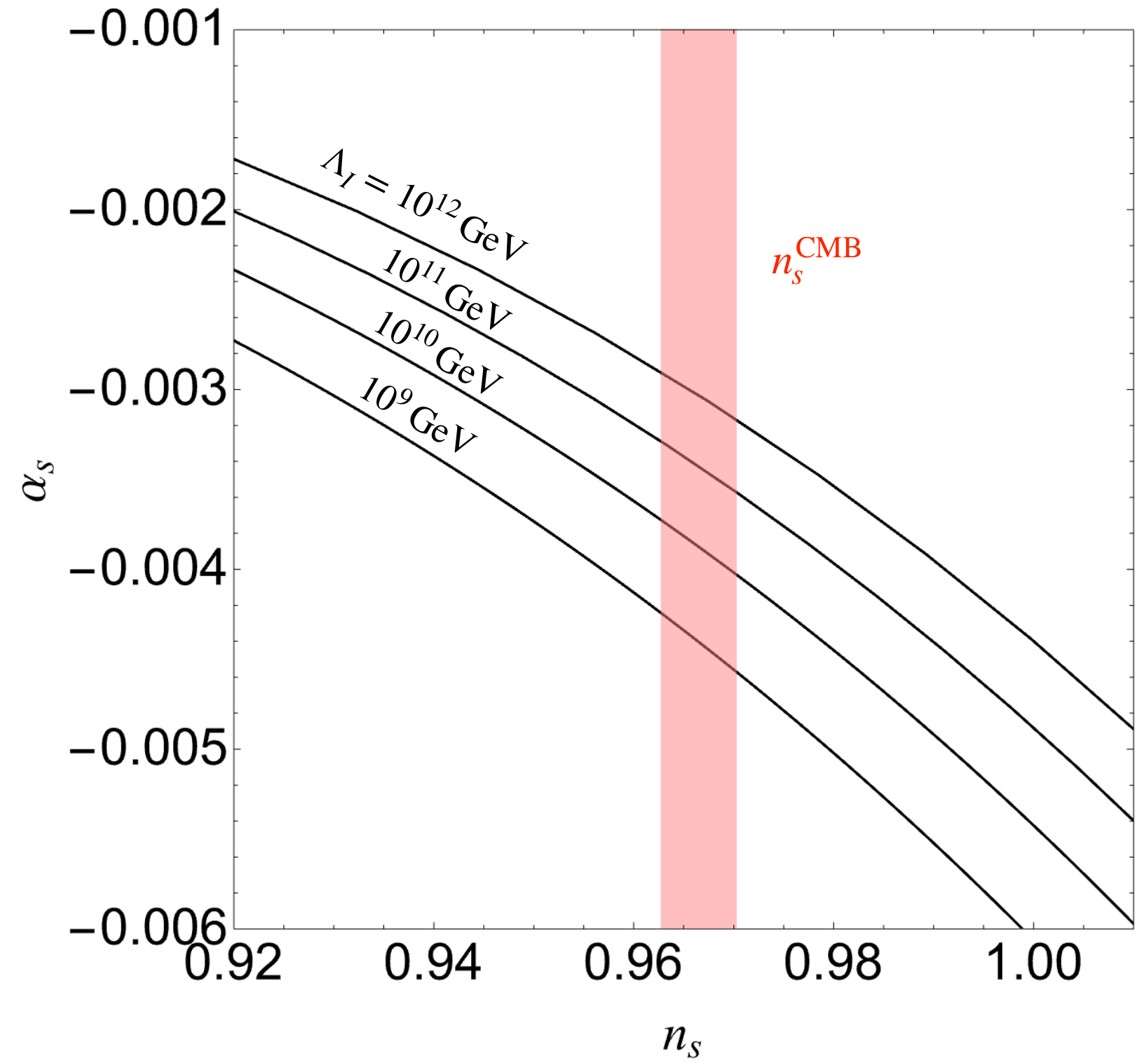}
\end{center}
\caption{ 
The prediction of the spectral index and its running with $\Delta_{\cal R}^2[\f_*]= \Delta^2_{{\cal R},\rm CMB}[k_*]$, by varying one-dimensional combination of $\d \x,\d M^{-2}$. $\Lambda_I=10^{12},10^{11},10^{10}, 10^9\GEV$ from top to bottom with fixed $y_t=0.5$.  The CMB measurement is also shown in the pink region. 
}
\label{fig:ns}
\end{figure}

We have shown that as the leading order of UV corrections, we can use \Eq{dEF}. 
With it, $|V_{EF,\f\f}|$ is slightly suppressed at the horizon exit, $\f$ at which is also modified, to enhance $n_s$ while keeping $\Delta_{\cal R}^2[\f_*]=\D^2_{{\cal R},\rm CMB}[k_*]$ intact. 
By using the full potential of \Eq{eft}, I estimated $n_s$ and the running by including the third order of slow-roll expansions \cite{Gong:2001he}. 
The predictions and parameter relations are shown in Fig.\ref{fig:4} by fixing $\D^2_{\cal R}[\f_*]=\D^2_{{\cal R},\rm CMB}[k_*].$ 
We see that $\d \x /\x \sim \O(\L_I/M_{\rm pl})$ can explain $n_s$. $1/M^2$ is at the value that there is an almost inflection point around the horizon exit, i.e. $V_{EF,h}\approx 0, V_{EF,hh}\approx 0$. 
This scenario, therefore, can explain the CMB data. 
In addition, we have the prediction, 
\beq
r\approx 1.3\times 10^{-27} \(\frac{\Lambda_I}{10^{11}\GEV}\)^4, ~~~-\a_s= (3\text{-}4)\times 10^{-3}
\eeq
i.e.,  tensor to scalar ratio is quite small that should not be reached in the near future but the running of the spectrum index, $\a_s\equiv {d n_s\o d\log k_*}$, is rather sizable (see Fig.\ref{fig:ns}) when $n_s=n_s^{\rm CMB}$ and can be probed in the future CMB experiments such as CMB-S4~\cite{CMB-S4:2016ple} and SPHEREx~\cite{Dore:2014cca}, who can measure the $\a_s$ at the precision of $10^{-3}$\cite{Munoz:2016owz} together with DESI~\cite{DESI:2013agm}, WFIRST~\cite{https://doi.org/10.48550/arxiv.1503.03757}, or/and SKA~\cite{CosmologySWG:2015ysq}. It may be worth mentioning that the negative $\a_s$ is slightly favored by the recent Planck data~\cite{Planck:2018vyg}.
The running of running of the spectral index, $\beta_s=\O(\eta \a_s)$, is more suppressed $|\beta_s|<\O(10^{-4})$. 

Although I used the full potential of \Eq{eft}, I have checked that the $\O(h^{-6})$ terms in the large $h$ expansion in \Eq{VEF} 
and $V_0 \O(h^4/M^4)$ terms in the small $h$ expansion 
are irrelevant in estimating the
relations between $\a_s$ and $n_s$ thanks to the intermediate range of $h$ for inflation.  Therefore our prediction is robust under the higher order corrections if the potential can be expanded as \Eqs{VEF} and \eq{dEF}.

\section{Discussions, future directions and conclusions}
\lac{4}
\paragraph{Successful inflation is close to the criticality between eternal and non-eternal inflation}
To have successful inflation, we need serious fine-tuning among the parameters, as conventional low-scale inflation does.\footnote{This tuning may be explained anthropically.} Here, I would like to show that requiring the successful inflation, i.e., with the consistent $\Delta^2_{\cal R}$ and $n_s$ with $\L_I=10^{9-12}\GEV$, the system happens to be around the criticality between the eternal and non-eternal inflation. 
The tuning of the potential may be explained by some mechanism relevant to the criticality. 
To see this, in the top panel of Fig.\,\ref{fig:pot}, I displayed the potential behavior in  $h/M_{\rm pl}$ -$d \log V_{EF}/d\log h$ plane with $\d \x=\d \x_{\rm sample}\equiv-3.888888889\times 10^{-7} \x_0, M^{-2}=4.137259718\times 10^{-6} M_{\rm pl}^{-2}$ (See \Eq{xi} for  $\x_0$).\footnote{I emphasize that the parameter here for the successful inflation is very sensitive to the SM parameters or approximations for the Higgs potential. Thus they are just provided for the sake of reproduction of the results, in which case, we have to use $y_t$, $M_{\rm pl}$, etc., the same as what I used. 
A different choice of the SM parameters does not change our main conclusions or the near criticality feature, but $\x, M^{-2}$ for the successful inflation would be different. Indeed, I have checked that various features are unchanged with different parameter sets. } The parameter set gives $n_s\approx 0.967, \alpha_s\approx-0.0033,\beta_s\approx-0.000048, \AND r\approx 1.3\times10^{-27}$, which are consistent with the CMB observation and can be tested in the future from $\a_s$.  The end of inflation and the horizon exit is shown by a blue circle and red star, respectively, and thus the internal dashed-purple line denotes the regime for the slow roll for the observable Universe. $d V_{EF}/d h$ becomes larger again for larger $h$. In this case, during the one Hubble time, the classical motion for the canonical field,
 $|\D_{\rm classical} \f| \sim |d V_{EF}/d\f| \times (3H_{\rm inf}^{2})^{-1}=|d V_{EF}/dh Z^{-1/2}|\times (3H_{\rm inf}^{2})^{-1}$,  is much larger than the quantum diffusion, $\D_{\rm quantum} \f \sim H_{\rm inf}/2\pi$ at any field value with $h<M,$ since $|\D_{\rm quantum}| \f\sim 10\GEV$ which is much smaller than $|\D_{\rm classical}| \f \gtrsim 10^{5-6}\GEV$. 
 The total e-folding in the range $h \lesssim M$, equivalently $\f \lesssim 10 M_{\rm pl}/\sqrt{\x}$, is at most $\sim 100$, c.f. $\sim 43$ for the observable Universe. 
 Therefore eternal inflation does not occur with $h\ll M$.  
  
In the bottom panel of Fig.\ref{fig:pot}, on the other hand, I decreased $\d \x (<0)$ by $5\times 10^{-6} \%$, i.e.,  increased $\x$ by $2\times 10^{-12}\%$, from the value in the top panel. 
  In this case, we have $\Delta_{\cal R}^2\sim 10 \Delta^2_{{\cal R},\rm CMB}$ at the horizon exit, and it cannot explain the CMB data, meaning that to have the successful inflation in the top panel, we need a serious tuning of the parameters. 
 On the other hand, the interesting thing is that this very tiny modification changes the potential shape qualitatively. 
The red curve shows that $V_{EF,h}$ is negative, meaning that we have two points where $V_{EF,h}=0$: a hilltop and a false vacuum from left to right. 
At the both two points the slow-roll conditions are satisfied and $|\D\f_{\rm quantum}|>|\D \f_{\rm classical}|=0$. Therefore eternal inflation can take place. As long as we explain EW vacuum stability with large $\x$, the false vacuum, and thus eternal inflation regime, still exist with $\d \x < (1+\O(10^{-6})\%)\d \x_{\rm sample}(<0).$ This is clear because the $1/M^2(>0)$ term lifts the potential at large $h$, while a sufficiently smaller $\x$ than $\x_0$ would lead the potential hilltop in the regime $h>M_{\rm pl}/\sqrt{\x_0}.$
The behavior of near criticality is found in the case $\L_I=10^{9,10,12}\GEV$ when $n_s = 0.96-0.97$.  
I conclude that successful inflation lies in the regime very close to the criticality between the Higgs potentials that lead to eternal and non-eternal inflation.  
Since $\xi_0=\O(10^{14})$ and the criticality of the potential is within the change of  $\xi$ by $10^{-12}\%$, 
$\D \x=\O(1)$ can reach the criticality.\footnote{From the degeneracy in $\k$ in \Eq{kappa}, this is also equivalent to the change of $\mu_H^2$ by fixing $\x$. Interestingly, we note that the criticality can reach with a change of $\mu^2_H$ by $\D \mu_H^2 \sim (100\GEV)^2$ for $\L_I\sim 10^{11}\GEV$. The change by the weak scale can change the potential to have a false vacuum.
This is due to the interesting coincidence $\L_I^2/M_{\rm pl} \sim \mu_H$ from the SM parameter measurement. 
Indeed, there is another coincidence for the QCD scale $\L_{{\rm QCD}JF}[h_{\rm hilltop}] \sim 100\GEV$~\cite{Matsui:2020wfx}. 
This may be relevant to some new principles and mechanisms for the origin of those scales.  
} 
By introducing another light field to slowly change the relevant parameters during inflation, the ``eternal inflation" end when the parameters are around the criticality. Then most Universe after inflation and reheating would have the parameters around the criticality (c.f. \cite{Yin:2021uus}).\footnote{This mechanism for alleviating the tuning may be generic in inflection-point inflation in which by explaining $n_s$ the inflation is non-eternal~\cite{Takahashi:2019qmh}.}
\begin{figure}[t!]
\begin{center}  
\includegraphics[width=60mm]{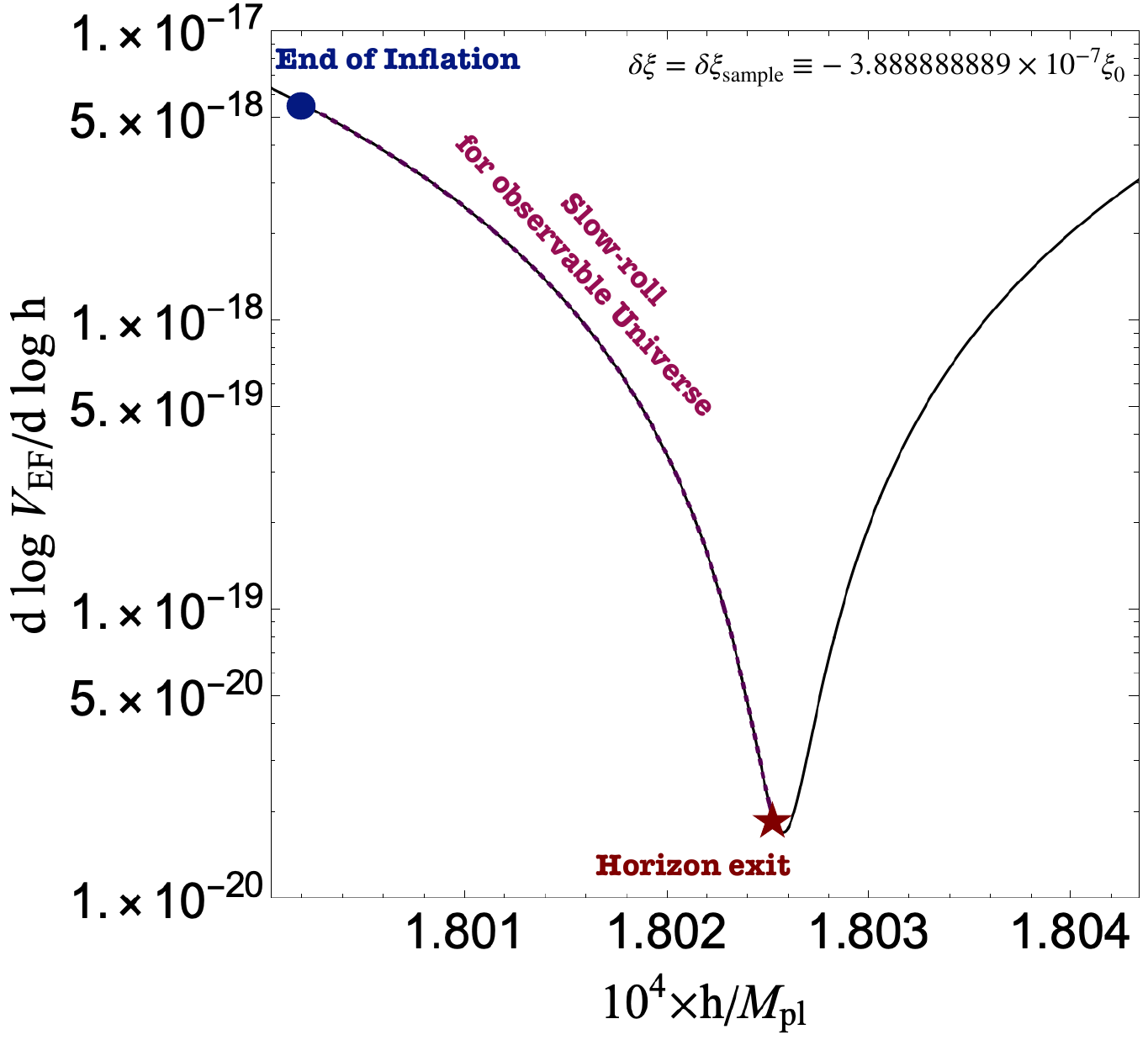}
\includegraphics[width=55mm]{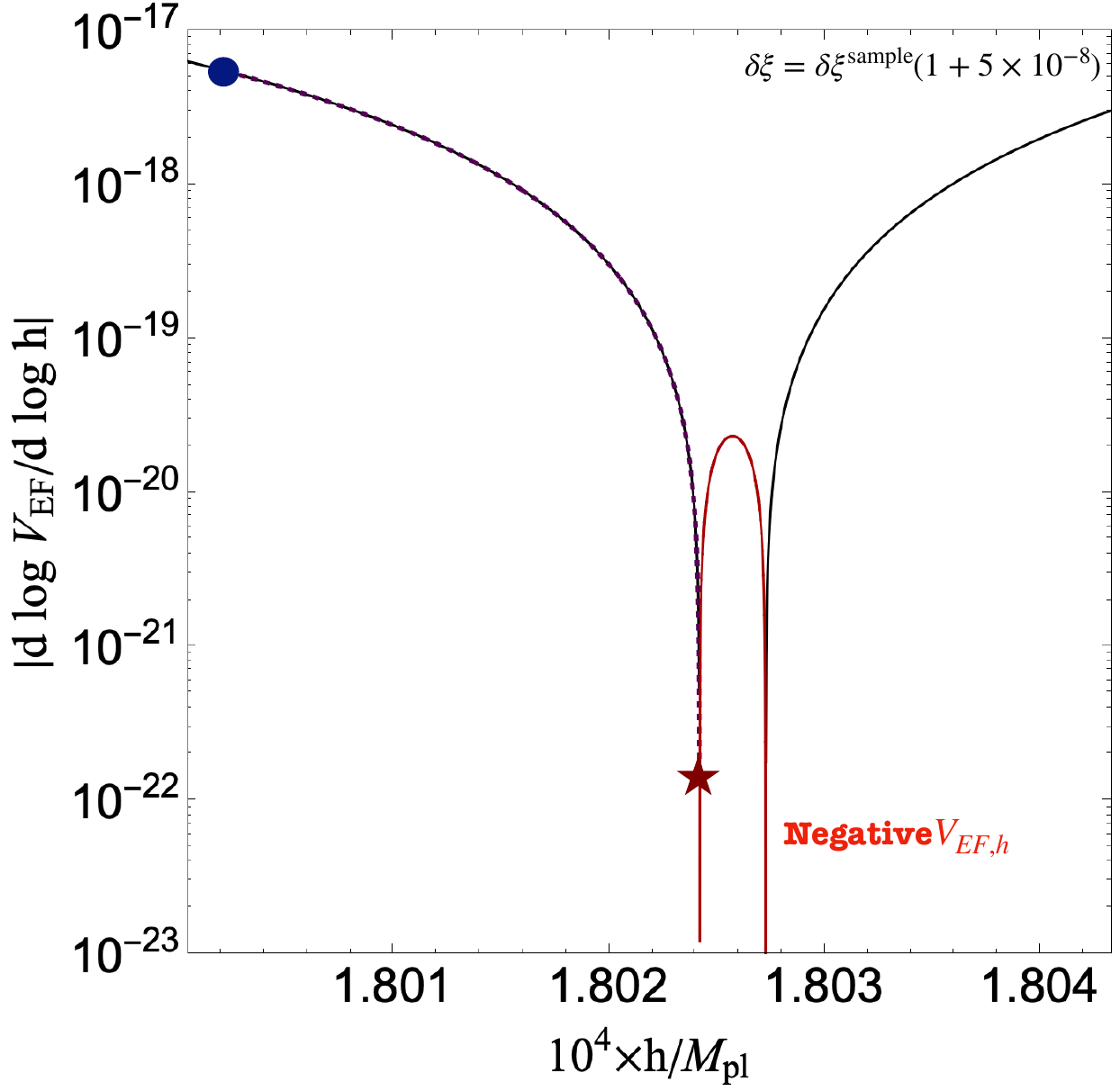}
\end{center}
\caption{ 
$d \log V_{EF}/d\log h$ vs $h/M_{\rm pl}$. In both panel, $\L_I=10^{11}\GEV, y_t=0.5, \m_H^2=(125\GEV)^2/2, \AND M^{-2}=4.137259718\times 10^{-6} M_{\rm pl}^{-2}.$
The difference is that, in the upper panel, I take $\x=\d \x^{\rm sample}\equiv -3.888888889\times 10^{-7} \x_0$, which explain the CMB data well, while, in the lower panel, $\x=\(1+5\times 10^{-8}\) \d \x^{\rm sample}$, which cannot explain the CMB data. 
}
\label{fig:pot}
\end{figure}

\paragraph{Dark matter production from inflationary fluctuation}
It was pointed out that if the weak-scale inflation happens due to the Higgs field excursion, the QCD axion with a decay constant around the Planck scale can be the dominant dark matter due to the enhanced QCD scale, $\sim 100\GEV$, by the $\sim \L_I$ scale SM quarks and renormalization group running,  and due to the stochastic behavior during the eternal inflation~\cite{Matsui:2020wfx}. Weak scale (Palatini) Higgs inflation does not allow this scenario because the inflation is not so long, and even if it is long due to some modification, the QCD scale is $\sim 0.1\GEV \frac{10^{3}}{\Omega}$ due to the large $\Omega$ during the inflation. The stochastic scenario for the QCD axion should not work~\cite{Graham:2018jyp, Guth:2018hsa}, and, naturally, the decay constant is around $10^{11}\GEV$ to explain the axion dark matter abundance. It may be interesting to study the hybrid inflation with the Peccei-Quinn and SM-like Higgs field. 
Other light axionic or scalar dark matter can be successfully produced~\cite{Ho:2019ayl,Takahashi:2019pqf, Nakagawa:2020eeg}. 
The weak-scale inflation, on the other hand, is known to alleviate the cosmological moduli problem~\cite{Randall:1994fr,Ho:2019ayl}.
In any case, we have to be careful of the change of the masses of the scalars in estimating the fluctuation in the inflationary era. This suppression of the dark matter mass during inflation should alleviate the isocurvature problem for relatively heavy dark matter (c.f. \cite{Ema:2018ucl}) due to the larger stochastic scalar amplitude. 
 Another interesting production of the light-dark matter should be from the decay of the ``inflaton", which has the smaller mass than the reheating temperature~\cite{Moroi:2020has,Moroi:2020bkq}. In particular, the Higgs inflaton couples to all possible dark matter through the very large non-minimal coupling. Thus, the production of dark matter could be an interesting future topic (c.f.~\cite{Li:2021fao,Li:2022ugn}).

\paragraph{Neutrino mass and leptogenesis}
Let us consider the seesaw mechanism to generate the neutrino mass~\cite{Minkowski:1977sc, Yanagida:1979as, Glashow:1979nm, GellMann:1980vs, Mohapatra:1979ia} since the $h$ value during inflation can be around or beyond the seesaw scale. 
During inflation, the right-handed neutrino (RHN) masses, whose Majorana components are suppressed by $\Omega$, are dominantly induced by the Higgs-neutrino-RHN Yukawa interaction.  
By estimating the Coleman-Weinberg potential from the Higgs-neutrino-RHN system  we can find that the contribution can be regarded as the redefinitions of $\k$ in \Eqs{VEF} in the large $h$ limit while the $\O(h^{-4})$ term rarely change as long as the RHN Yukawa coupling is much smaller than $y_t$. 
Our conclusions in the main part do not change.

This inflation scenario which has a reheating temperature, $T_R = 10^{8-11}\GEV,$ can have a successful leptogenesis~\cite{Fukugita:1986hr,Pilaftsis:1997dr,Buchmuller:1997yu,Akhmedov:1998qx,Asaka:2005pn} (see also 
Refs.\,\cite{Buchmuller:2005eh, Davidson:2008bu}) when the RHNs are light enough after inflation. 

On the other hand, after the inflation, the RHNs may get heavier than the reheating temperature due to the smaller $\Omega$, and thus the thermal or direct production from the Higgs interaction to produce the RHN may be kinematically suppressed. 
In this case we can still have successful leptogenesis via the left-handed lepton oscillation~\cite{Hamada:2018epb} (See also Refs.\,\cite{Hamada:2016oft, Eijima:2019hey}). Indeed there can be an enhancement in the scenario because soon after inflation, we should have an over-dense system during the first periods of reheating~\cite{Eijima:2019hey}. 
\paragraph{Conclusions}
The precise SM parameter measurements and the renormalization group running suggest that there is a hilltop of the Higgs potential in the intermediate scale, and the EW vacuum is metastable. In this paper, it was pointed out that introducing a very large Higgs non-minimal coupling can make the EW vacuum the true one and, thus, absolutely stable. The same setup can drive slow-roll inflation, which enhances the density perturbation to the measured level thanks to the predicted hilltop in the SM. We have studied this minimal inflation scenario carefully and showed in the Palatini Higgs inflation that the robust prediction of the running of the spectral index of the scalar perturbation can be tested in future CMB experiments. It is also important to have a future collider to measure the SM parameters more precisely to clarify the Higgs potential shape in more detail.

\section*{Acknowledgments}
I thank Seong Chan Park for the useful comment. 
WY was supported by JSPS KAKENHI Grant Nos.  20H05851, 21K20364, 22K14029, and 22H01215.

\bibliography{references}

\end{document}